\documentclass[prl,twocolumn]{revtex4}%
\usepackage{amsmath}
\usepackage{amsfonts}
\usepackage{amssymb}
\usepackage{graphicx}%
\begin{document}


\title{An Infrared Spatial and Frequency Selective Metamaterial Perfect Absorber}

\author{Xianliang Liu$^1$, Tatiana Starr$^2$, Anthony Starr$^2$, and Willie J. Padilla$^1$$^\star$}
\affiliation{$^1$Department of Physics, Boston College, 140 Commonwealth Ave., Chestnut Hill, MA 02467, USA}
\affiliation{$^2$SensorMetrix, Inc., San Diego, California 92121, USA}

\email{willie.padilla@bc.edu}

\begin{abstract}

We demonstrate, for the first time, a spatially dependent metamaterial perfect absorber operating in the
infrared regime. We achieve an experimental absorption of 97$\%$ at a wavelength of 6.0
microns, and our results agree well with numerical full-wave simulations. By using two different metamaterial sublattices
we experimentally demonstrate a spatial and frequency varying absorption which may have many
relevant applications including hyperspectral sub-sampling imaging.

\end{abstract}

\maketitle

Since the first experimental demonstration of negative refractive index~\cite{Veselago68, Smith04, Smith00, Schultz01}, research into metamaterials has grown enormously. The ability of metamaterial to achieve nearly any electromagnetic response in nearly any frequency band suggests many exotic applications including invisibility cloaks~\cite{Schurig06} and perfect lenses~\cite{Pendry00, Fang05}. One current expanding field of research is that of metamaterial perfect absorbers (MPAs) due to their unique ability to achieve unity absorption with high efficiency~\cite{Landy08}. Here we present a computational and experimental study of an infrared metamaterial absorber which realizes 97$\%$ absorption at 6.0 micron wavelength. By using two different metamaterial sub-lattices consisting of a MPA and a near zero absorber, we experimentally demonstrate a spatial and frequency varying absorption which may have many relevant applications including hyperspectral sub-sampling imaging.

Metamaterials, through changes of the size and shape of subwavelength metallic elements, permit the amplitude and frequency tuning of both the electric [$\epsilon(\omega)$] and magnetic [$\mu(\omega)$] response of electromagnetic radiation. Although this electromagnetic response is resonant and narrow band, there have been demonstrations of exotic metamaterials operating in all relevant bands for frequencies below visible~\cite{Shalaev05,Linden04,Zhang05,Yen04,Bayindir02,Wiltshire01}. The resonant nature of metamaterials results in a strong focusing of the electric field within gaps of the structure, thus proving means to dynamically control the resonance frequency, phase, and amplitude~\cite{Chen08, Chen06, Chen09}. By using either passive or active external stimuli; temperature, electric field, optical, etc., dynamical MPAs may be explored and extended to much of the electromagnetic spectrum~\cite{Tao082, Tao08}. After first demonstration of the microwave MPA, many efforts have focused on extending highly absorbing metamaterials to smaller wavelengths. To-date most studies have been carried out in the terahertz regime and absorptions as high as 96.8 $\%$ have been achieved~\cite{Tao082}.

Recently there is great interest in single-pixel imaging, based on the theory of compressive sensing~\cite{Donoho06,Tao06,candes06}. Typically, experimental realization of these systems utilizes spatial light modulators (SLMs) for encoding random spatial patterns in the wave front of transmitted or reflected optical radiation~\cite{Gehm07, Chan208}. In the THz frequency regime metamaterials have shown promise to be used as SLMs in a transmission configuration~\cite{Chan09}. At visible wavelengths single pixel imagers may use liquid crystals or Micro-Electro-Mechanical Systems (MEMS) to manipulate the spatial profile of the beam. However at lower infrared frequencies liquid crystals are inefficient thus preventing their use. Further, for many applications, mechanical manipulation of the incoming image may not be feasible. An intriguing application of metamaterial absorbers would be their use as a SLM in a sub-sampling infrared imaging system. However, to-date, metamaterial absorbers have only been demonstrated in the microwave and THz frequency regimes.

\begin{figure}[!]
\begin{center}
\includegraphics[ width=2.75in,keepaspectratio=true]%
{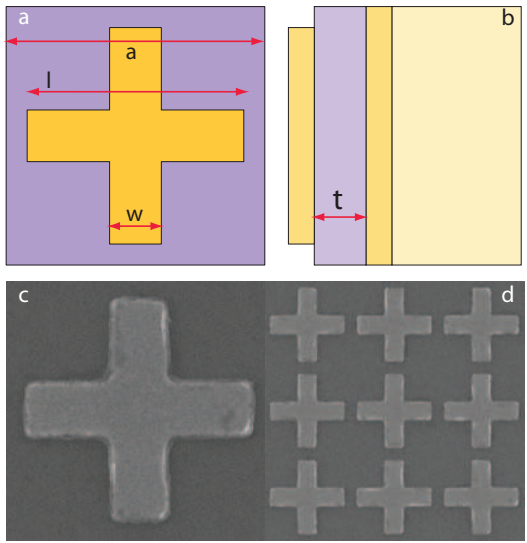}%
\caption{\textbf{Design and fabrication of the infrared metamaterial absorber.} a, Schematic of a MPA unit cell and its optimized dimensions are a = 2, l = 1.7, w = 0.4, t = 0.2 in micrometer. b, Side view of the unit cell. c, Scanning Electron Microscope (SEM) image of one unit cell. d, SEM image of a periodically patterned array.}
\label{Fig1}%
\end{center}
\end{figure}

The schematic of a single unit cell of an infrared MPA is shown in Fig. \ref{Fig1}(a) and (b), and consists of two metallic elements; a cross shaped resonator and ground plane. We space these metallic elements apart by virtue of a dielectric layer. The cross resonator is a type of electric ring resonator (ERR)\cite{Schurig06, padilla07} and couples strongly to uniform electric fields, and negligibly to magnetic fields. However, by paring the ERR with a metallic ground plane, the magnetic component of light couples to both the center section of the cross resonator and the ground plane, thus generating antiparallel currents resulting in resonant response. The magnetic response can therefore be tuned independent of the electric, by changing the geometry of the ERR and the distance between elements. By tuning the amplitude and frequency location of the electric and magnetic resonances it is possible to match the impedance $Z(\omega)=\sqrt{\mu(\omega)/\epsilon(\omega)}$ of the MPA to free space, thus minimizing the reflectance at a specific frequency. The metallic ground plane is thicker than the penetration depth of light in the IR range and thus transmission of the MPA is zero. These two conditions thus lead to narrowband high absorption.

\begin{figure}[!]
\begin{center}
\includegraphics[ width=3in,keepaspectratio=true]%
{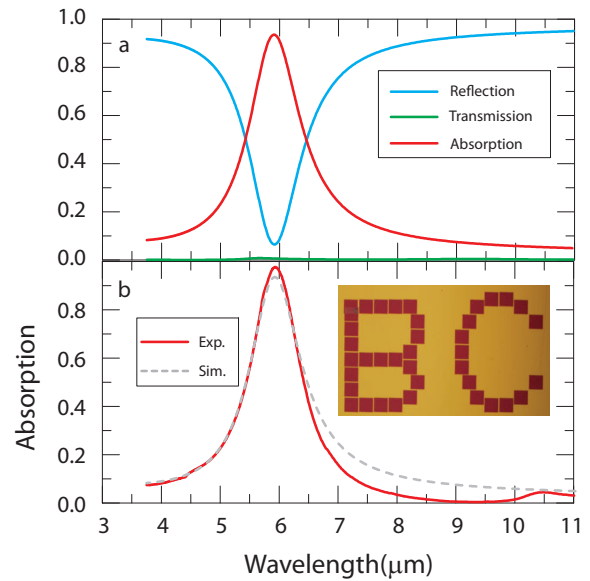}%
\caption{\textbf{Simulated and experimental performance of the infrared metamaterial absorber} a, Numerical simulations of the electromagnetic response: Reflection (blue), Transmission (green) and Absorption (red). The maximum absorption at 6 $\mu$m is 97$\%$. b, A comparison between the experimental absorption (red) and simulated absorption (dashed grey). Inset to (b) shows an optical microscope image of the fabricated spatially dependent metamaterial absorber.}
\label{Fig2}%
\end{center}
\end{figure}

The optimized structure presented in Fig. \ref{Fig1} was obtained through computer simulations using the commercial program CST Microwave Studio 2009. The time domain solver was utilized and metallic portions of the metamaterial absorber were modeled as gold using a Drude model with a plasma frequency $\omega$$_p$=2$\pi$$\times$2.175$\times$10$^{3}$THz  and collision frequency $\omega$$_c$=2$\pi$$\times$6.5THz~\cite{Ordal83}. We use Al$_2$O$_3$ as the dielectric spacer and simulate the dielectric constant and loss tangent as 2.75 and 0.02 respectively. The transmission T($\omega$)=$|$S$_{21}(\omega)|^2$ and reflection R($\omega$)=$|$S$_{11}(\omega)|^2$ were obtained from S-parameter simulations with appropriate boundary conditions to approximate a TEM wave incident on the structure with both electric and magnetic field vectors lying in plane and the wavevector perpendicular. The frequency dependent absorption was calculated as A($\omega$)=1-R($\omega$)-T($\omega$) where, as expected for the present design, S$_{21}$ is zero across the entire frequency range due to the metallic ground plane. The optimized structure, Fig. \ref{Fig1}(a), achieves a simulated maximum absorption at 6.0$\mu$m and has the geometrical parameters, (in microns), of: a = 2.0, l = 1.7, w = 0.4, t = 0.2, and the thickness of both metallizations was 0.1 $\mu$m. The simulated R($\omega$), T($\omega$), and A($\omega$) are shown in Fig. \ref{Fig2}(a).

Simulations indicate the feasibility of realizing an infrared MPA, and we now thus turn toward experimental demonstration. Fabrication begins with E-beam deposition of a 100 nm thick layer of gold on a silicon substrate. This is followed by Atomic Layer Deposition (ALD) of a 200 nm thick layer of Al$_2$O$_3$. Resist 950 PMMA-A (MicroChem) was then spin coated on top and patterned using E-beam lithography. Another 100nm thick layer of gold was E-beam evaporated followed by liftoff in acetone. Fig. \ref{Fig1}(c) displays a Scanning Electron Microscope (SEM) image of a single unit cell of the fabricated sample and Fig. \ref{Fig1}(d) shows a larger field of view. The total lateral size of the sample was 140$\mu$m $\times$ 140$\mu$m.

Transmission and reflection were characterized from a wavelength of 3$\mu$m to 12$\mu$m using a Fourier-transform infrared spectrometer combined with an infrared microscope (liquid-N$_2$-cooled MCT detector, 15$\times$ cassegrain objective lens and KBr beam splitter). The reflection was measured at an incident angle of 20 degrees and transmission was measured at normal incidence. The measured reflection spectra are normalized with respect to a gold mirror while the transmission spectrum is normalized with respect to an open aperture. The measured transmission and reflection are then used to calculate the absorption.

\begin{figure*}[!]
\begin{center}
\includegraphics[ width=5.5in,keepaspectratio=true]%
{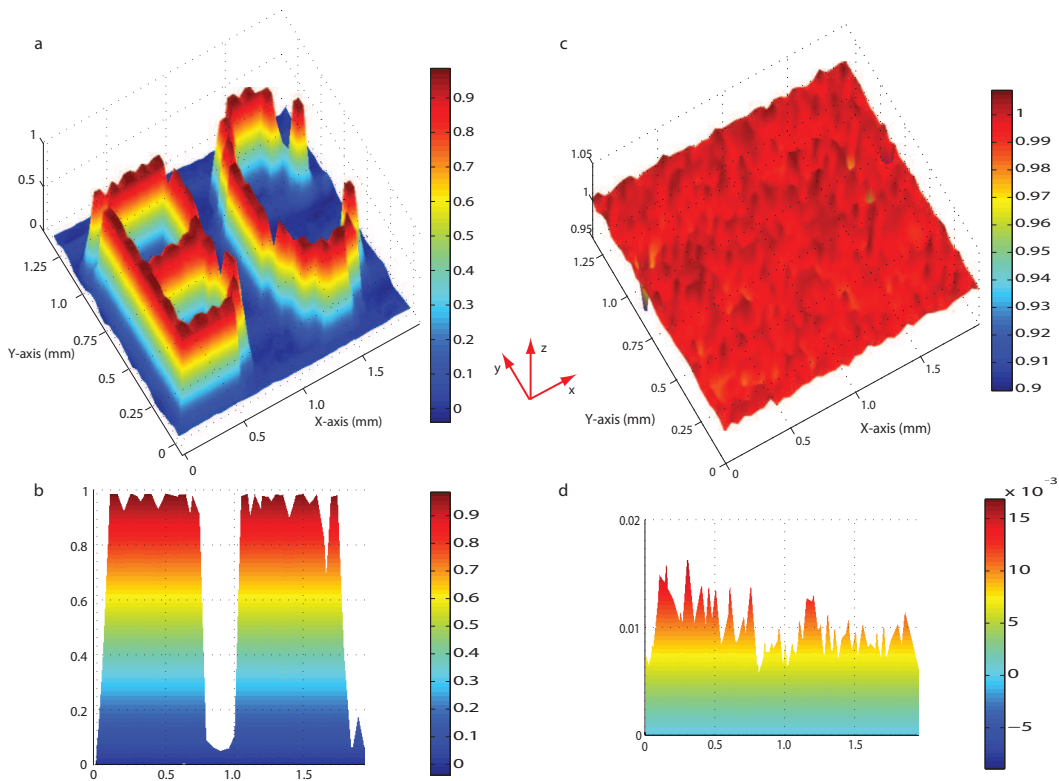}%
\caption{\textbf{Experimental three dimensional images of the absorption and reflection of the graded metamaterial absorber.} a,  Spatially dependent absorption image at a wavelength of 6 $\mu$m (plotted to 100$\%$ absorption). b, x-z view of the absorption at 6 $\mu$m. c, Image of the reflection at a wavelength of 10 $\mu$m. d, x-z view of the reflection at 10 $\mu$m.}
\label{Fig3}%
\end{center}
\end{figure*}

The experimental absorption is shown in Fig. \ref{Fig2}(b) as the red solid curve. As shown in Fig. \ref{Fig2}(a) the simulated transmission is effectively 0 across the entire range and a resonance occurs in the simulated reflectance at 6 $\mu$m where R=0.03, thus yielding a maximum in the simulated absorption. The experimental absorption has a maximum at 6.0 microns with a value of 97$\%$, with relatively low values to both lower and higher frequencies. The absorption has a full width half maximum of 1.0 $\mu$m and, at wavelengths of 4 $\mu$m and 10 $\mu$m A($\omega$), has fallen off to 10$\%$ and 1$\%$ respectively. We achieve good agreement between experimental and simulated (dashed grey curve) absorptivities, as is evident from Fig. \ref{Fig2}(b).

In order to demonstrate the versatility of the MPA and its potential for applications, we combined our highly absorbing metamaterial, which we now term Unit Cell A (UCA), with an extremely low absorbing metamaterial, Unit Cell B (UCB). This is a type of metamaterial consisting of different sub-lattices, of which both bipartite,~\cite{yuan} and tripartite~\cite{bingham} variations have been demonstrated. Unit cell B is of identical lateral proportions as UCA, i.e. 2$\mu$m$\times$2$\mu$m, and consists of the same metallic ground plane and Al$_2$O$_3$ dielectric with identical thicknesses. We use each of these two unit cells to form a pattern in the xy-plane, i.e. we form a spatially selective or `graded' perfect absorber. The inset to Fig. \ref{Fig2}(b) displays an optical microscope image of just such a design, forming the two letters `B' and `C'. The pattern of UCA consisted of 39 writing fields, each of dimensions 140$\mu$m $\times$ 140$\mu$m, where each contains nearly 5000 metamaterial unit cells. To avoid overlap we use a 4 $\mu$m space between neighboring UCA writing fields.

We characterized the graded metamaterial absorber using the same experimental setup as described earlier. A computer controlled linear stage moved automatically in the xy-plane and we collected both R and T from wavelengths of 20$\mu$m to 600 nm. Reflection spectra is normalized with respect to a Au mirror with the same beam size. A spatial resolution of 50$\mu$m was used thus creating an image with 40 pixels along the x-axis and 28 along the y-axis. A total of 16,000 interferrogram points were collected for each pixel. A two-dimensional hyperspectral image is acquired, with a total data cube size of 1.8$\times$10$^7$, thus permitting investigation of both the frequency and spatial dependence of the sample.

We display 3D images of the data for a wavelength of 6 $\mu$m, shown in  Fig. \ref{Fig3}(a)(b). The image is displayed as (x,y,A) and thus the \^{x} and \^{y} axis indicate the dimensions of the pattern and the \^{z} axis displays the absorption for each pixel. Clearly, as demonstrated in Fig. \ref{Fig3}(a), at the resonance frequency of $\lambda$ = 6 $\mu$m, the patterned metamaterial (Unit Cell A) achieves near unity absorption. Fig. \ref{Fig3}(b) shows the x-z plane from which it can be directly observed that absorption is near zero for UCB. Thus by using bipartite unit cells with perfect absorption and with near zero absorption, we have maximized contrast of the image.

In order to show the frequency dependence of the spatially selective MPA, in Fig. \ref{Fig3}(c)(d) we plot 3D images for a wavelength of $\lambda$=10 $\mu$m. Data is in the form (x,y,R), where R is the reflectance, for Fig. \ref{Fig3}(c) and (x,y,A) for Fig. \ref{Fig3}(d). As can be observed, the reflectance is close to unity over the xy-plane and, correspondingly, the absorption is near zero. A perfect metal at these wavelengths would give 100$\%$ reflectance and zero absorption. In contrast to both the optical photograph (Fig. \ref{Fig2}(b)), and the image at 6$\mu$m (Fig. \ref{Fig3}(a)), at $\lambda$=10 $\mu$m the MPA is not visible, i.e. indistinguishable from a perfect metal.

In conclusion, we have designed, fabricated and characterized a mid-infrared metamaterial perfect absorber, which achieves an experimental absorption of 97$\%$ at a wavelength of 6.0$\mu$m. We further patterned the MPA to demonstrate the ability of achieving both a frequency and spatially selective absorption. Since one may specify a particular spatial pattern to be sent to a detector and the other component to be strongly absorbed, we have the ability to spatially modulate signals with large dynamic range. Dynamical range as large as 40 db is achieved for our MPA without optimization which is comparable to existing digital cameras~\cite{Battiato03}. These result suggest an alternative way to achieve spatial light modulation, information coding, and single pixel imaging. Interestingly, by combining the MPA with dynamical control~\cite{Chen06, Chen08}, the option of performing real time sub-sample imaging, and hyperspectral sub-sample imaging, becomes possible. One advantage of dynamical metamaterials is that they may be fabricated with no moving parts,\cite{Chen06} i.e. they are all solid-state. Due to the scalability of metamaterials, the concept of a spatially selective MPA suggests the ability of constructing SLMs which may operate at any sub-visible frequency. The above listed advantages possible with metamaterials suggests exotic applications - in particular hyperspectral single pixel imaging cameras, which may alleviate the typically incurred trade-offs between cost, speed, signal-to-noise ratio, spectral resolution, and spatial resolution.

We Acknowledge Tom Tague and Fred R. Morris from Bruker Optics for their help on sample characterization. This research was funded by the Office of Naval Research under U.S. Navy Contract N00014-09-M-0290.

\end{document}